\documentclass[a4paper]{jpconf}
\usepackage{graphicx}
\begin{document}
\title{Neutrinoless Double Beta Decay with SNO+}

\author{J Hartnell for the SNO+ collaboration}

\address{Department of Physics and
    Astronomy, University of Sussex, Brighton. BN1 9QH\@. United
    Kingdom.}

\ead{j.j.hartnell@sussex.ac.uk}

\begin{abstract}
SNO+ will search for neutrinoless double beta decay by loading 780
tonnes of linear alkylbenzene liquid scintillator with O(tonne) of
neodymium. Using natural Nd at 0.1\% loading will provide 43.7~kg of
$^{150}$Nd given its 5.6\% abundance and allow the experiment to reach
a sensitivity to the effective neutrino mass of 100-200~meV at 90\%
C.L in a 3 year run. The SNO+ detector has ultra low backgrounds with
7000 tonnes of water shielding and self-shielding of the
scintillator. Distillation and several other purification techniques
will be used with the aim of achieving Borexino levels of
backgrounds. The experiment is fully funded and data taking with
light-water will commence in 2012 with scintillator data following in
2013.
\end{abstract}

\section{Introduction}
The use of the world's largest liquid scintillator detectors in the
search for neutrinoless double beta decay is a promising development
that has come to the fore in the last few
years~\cite{Chen:2005yi,Mitsui:2011zz}. These experiments push the
trade-off between having a large mass of available isotope but lower
energy resolution into a new regime. The re-use of the infrastructure
from the SNO experiment~\cite{Boger:1999bb} provides a cost-effective
and ultra low-background detector to search for neutrinoless double
beta decay. The first phase of the SNO+ experiment~\cite{Chen:2005yi}
will load about a tonne of natural neodymium (5.6\% $^{150}$Nd) into
the scintillator. Future possibilities include neodymium enriched in
$^{150}$Nd or different isotopes that can be loaded at higher
concentrations without significantly affecting the scintillator light
output. While not covered here, a summary of the diverse range of
other physics that will be probed by SNO+ including solar neutrinos,
geo and reactor antineutrinos, supernovae can be found
in~\cite{lozza}.

\section{Detector}
The SNO+ detector is located 2~km underground (6000~m.w.e.) at Vale's
Creighton nickel mine near Sudbury, Ontario, Canada. The expansion of
the scientific facility with the creation of SNOLAB has produced the
world's deepest operating scientific laboratory, of which the SNO+
experiment is part.

A barrel-shaped cavity 34~m high and 22~m wide filled with ultra pure
water provides the space for the detector. A 17.8~m diameter geodesic
stainless steel frame is suspended from the cavity deck and forms the
PMT Support Structure (PSUP) that holds the approximately 9500 8-inch
PMTs. Within the PSUP is a 12~m diameter 5.5~cm thick acrylic vessel
(AV) with an internal volume of 907~m$^3$. The primary change in the
conversion of the SNO detector into SNO+ is the replacement of the
heavy water in the AV with linear alkylbenzene (LAB) scintillator. The
LAB with a density of about 0.86~g/cm$^3$ will be buoyant in the water
of the cavity and so a rope-net system that passes over the top of the
AV and anchors it to the floor is being installed.

The PMTs are offset from the scintillator in the AV by about 3~m and
the intermediate water volume provides 1700 tonnes of
shielding. Outside the PSUP there are a further 5700 tonnes of water
that shield the detector from external backgrounds in the rock. The
walls and floor of the cavity itself are lined with Urylon, which is
expected to attenuate radon emanating from the walls by a factor of
nearly 10$^7$~\cite{Boger:1999bb}. The Urylon liner on the floor of
the cavity was successfully replaced in 2011 and now incorporates the
anchors for the new AV hold-down system.

A new trigger and DAQ system has been developed for SNO+ due to the
factor of about 45 more light expected from the scintillator compared
to the Cherenkov light in the heavy water. A new radon-tight universal
interface and cover gas system will be installed in 2012.

\subsection{Scintillator Systems and Purification}

The scintillator for SNO+ comprises of 780 tonnes of linear
alkylbenzene (LAB) with 2g/L of 2,5-diphenyloxazole (PPO) as a
wavelength shifter~\cite{ford}. LAB was selected due to its
compatibility with acrylic, high flash point, good light yield, low
toxicity and good commercial availability. An extensive set of
scintillator purification systems are under construction with the aim
of reaching the purity levels (10$^{-17}$ g/g of $^{238}$U and
$^{232}$Th chain activities) achieved by the Borexino
experiment~\cite{borexino}. The approach of Borexino is being followed
for SNO+, in particular with regard to the cleanliness and vacuum leak
tightness.

Multi-stage vacuum distillation of the LAB will be performed, prior to
adding it to the detector. The low volatility of the heavy metal
contaminants enable an extremely pure product to be produced. The
distillation system will operate at 19 L/minute allowing the detector
to be completely filled in 1-2 months. A concentrated PPO solution
will be flash distilled under vacuum at a higher temperature; mixing
with the bulk of the LAB will be done before it reaches the
detector. A further purification step involves a steam-stripping
process to remove the high volatility contaminants of the LAB which
include the noble gases (e.g. Rn, Ar, Kr) and oxygen.

While the detector is operational it will be possible to recirculate
the entire scintillator volume in 4 days (150 LPM) for in-situ
repurification. A rotating-stage water extraction column that exploits
the differences in solubility of contaminants in water vs. LAB will be
used to remove elements such as Ra, K and Bi. Columns loaded with
metal scavenger beads made by Johnson Matthey Plc (who acquired Reaxa
LTD) will be used to remove Pb, Ra, Bi, Ac, and Th. The steam
stripping process will also be used during recirculation, and the
scintillator will be subject to micro-filtration.

\subsection{Neodymium Loading and Purification}

The Nd will be loaded into the LAB at the O(0.1\%) level and
NdCl$_{3}$ is the starting point. The technique that will be used to
make a soluble Nd-compound is pH-controlled, solvent-solvent
extraction~\cite{bnlGd}. The carboxylic acid used will be 3,5,5
trimethyl hexanoic acid (TMHA) and it will be pre-purified using a
thin-film evaporator column. The NdCl$_3$ solution will be
pre-purified using the pH-adjustment co-precipitation
(self-scavenging) technique~\cite{bnlNdPur}, which exploits the large
difference in solubility between Nd and contaminants such as
Th. Laboratory tests have demonstrated the effectiveness of this
self-scavenging procedure. An important consideration for the
Nd-loading procedure is whether the purification systems used during
recirculation of the scintillator would remove the Nd and this has
been demonstrated not to be an issue for the water extraction and
stripping processes. The metal scavengers will not be used on the
Nd-loaded scintillator, but will be used as a polishing stage to
remove the last traces of Nd when required.

\section{Neodymium}

Neodymium is classed as a rare earth element but is actually widely
distributed in the Earth's crust, being as common as nickel and
copper. A common use for Nd is to make the high strength magnets
found, for example, in hybrid cars, computer hard disks, wind turbines
and the headphones for personal stereos. Natural Nd contains two
radioisotopes, $^{150}$Nd at 5.6\% natural abundance and $^{144}$Nd at
23.8\%. The $^{144}$Nd decays with a 1.91 MeV alpha and a half-life of
$2.3\times10^{15}$~years giving an activity of 7.5~kBq per tonne of
natural Nd. The double beta decay of $^{150}$Nd has been recently
measured by the NEMO-3 collaboration using 36.55~g of isotope to have
a half-life of $T_{1/2}^{2\nu}=9.11^{+0.25}_{-0.22}({\rm
  stat})\pm0.63({\rm
  syst})\times10^{18}$~years~\cite{Argyriades:2008pr}; the half-life
for neutrinoless double beta decay is found to be
$T_{1/2}^{0\nu}>1.8\times10^{22}$ years in the same experiment and is
currently the world's best limit.

\section{Sensitivity to Neutrinoless Double Beta Decay}

With 0.1\% loading SNO+ will use 0.78~tonnes of neodymium and contain
43.7~kg of $^{150}$Nd with no enrichment. Nd has the largest phase
space factor of all double beta decay isotopes and an endpoint of
3.37~MeV that places it above most backgrounds from natural
radioactivity.

\begin{figure}[ht]
%converted the gif file into png (or pdf) using Preview on my mac
%\includegraphics[width=1.0\textwidth,bb=1.0in 1.0in 7.5in 10in]{energysum_400hits_zoom.pdf}
\centerline{\includegraphics[width=90mm,height=60mm]{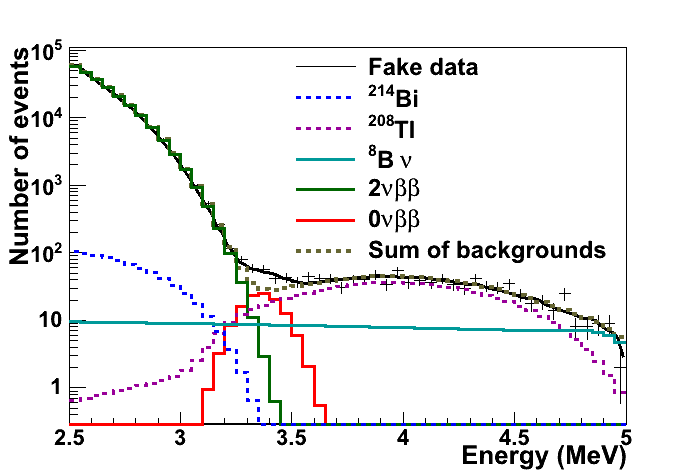}}
%24.56 × 16.66 cm, ratio=1.474
%With PNG or JPEG you should specify an explicit width or height rather
%than "scale", since bitmap images have no intrinsic size, nothing
%corresponding to Bounding Box information
\caption{Simulated event energy spectrum showing the signal
  corresponding to an effective neutrino mass of 350~meV for a 3 year
  run with 0.1\% Nd-loading.}
\label{energySpect}
\end{figure}

Backgrounds include the $2\nu\beta\beta$ from $^{150}$Nd, $^8$B solar
neutrinos, $^{208}$Tl and $^{214}$Bi. These backgrounds are shown in
Figure~\ref{energySpect} along with a $0\nu\beta\beta$ signal assuming
an effective neutrino mass of 350~meV. 
\begin{figure}[ht]
\centerline{\includegraphics[width=73mm,height=49mm]{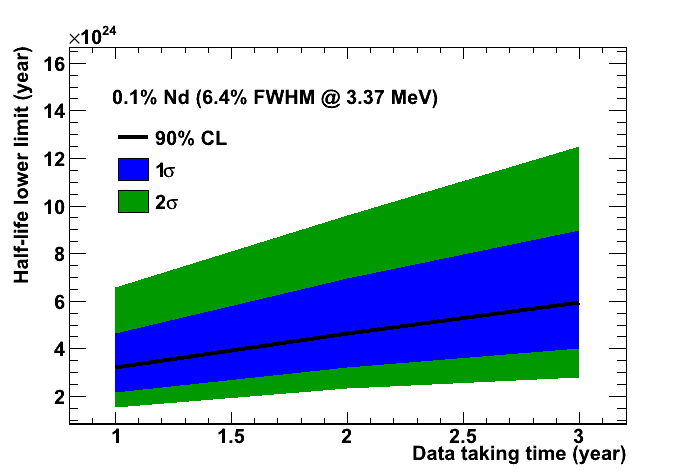}\hspace{0.1cm}\includegraphics[width=73mm,height=49mm]{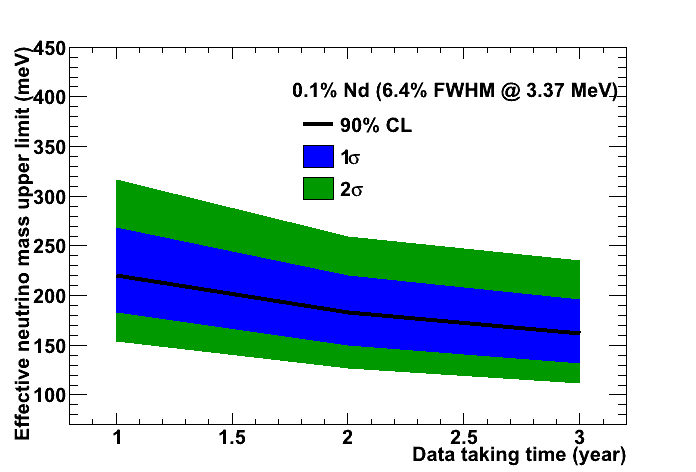}}
\caption{(Left) The evolution of the 90\% C.L. $0\nu\beta\beta$
  half-life sensitivity over a 3 year run with 0.1\% loading of
  Nd. (Right) The same calculation but showing the 90\% C.L. effective
  neutrino mass sensitivity.}
\label{sensitivity}
\end{figure}
Gaussian smearing was applied under the assumption that 400
photoelectrons are detected per MeV. $^{214}$Bi can be tagged, using
the decay of its daughter isotope $^{214}$Po in delayed coincidence,
with high efficiency and a conservative 90\% tagging efficiency is
assumed for all sensitivity calculations. Figure~\ref{sensitivity}
shows the evolution of the half-life and mass sensitivity over a 3
year run. The IBM-2 matrix element~\cite{Barea:2009zza} that includes
the effect of nuclear deformation is used to calculate the effective
neutrino mass sensitivity. A fiducial volume of 50\% of the AV, 80\%
detector livetime and backgrounds equivalent to those achieved in
Borexino were used in the sensitivity calculations. There are avenues
for improvement in all those areas and additionally, an optimisation
of the trade off between the Nd-loading concentration and scintillator
light output will be performed. Studies have shown that by increasing
the Nd-loading concentration to 0.3\% the light level drops by about a
factor of two but the half-life sensitivity increases by 70\%.

SNO+ will demonstrate the effectiveness of its technique with natural
Nd. However, the possibility of Nd isotopically enriched in $^{150}$Nd
is enticing and there are a few potential approaches including Atomic
Vapour Laser Isotope Separation (AVLIS) and high-temperature
centrifuge. Nd enriched to 80\% $^{150}$Nd would increase the
potential number of signal events by a factor of 16. As a simple
example, if the experiment were to run 16 times longer then a factor
of 2 increase in mass sensitivity would be expected. However, with
enriched Nd, backgrounds from $^8$B solar neutrinos and also
potentially $^{208}$Tl would stay constant indicating that more than a
factor of 2 increase in mass sensitivity could be possible. With an
eye to the future, a particular advantage of scintillator detectors
like SNO+ is that a variety of isotopes can be used, providing several
different potential opportunities.

\section{Schedule}
Data taking with the AV filled with light water will commence in 2012
and construction of the scintillator process systems will proceed in
parallel. In 2013 the scintillator that has passed through the
purification systems will gradually displace the water over a period
of a few months and scintillator-filled data taking will commence.

\section{Conclusion}
SNO+ is an ultra low background liquid scintillator detector located
at SNOLAB, Canada. The central acrylic vessel is surrounded by 7400
tonnes of water shielding and the scintillator also self-shields. The
scintillator will be purified initially by distillation and additional
in-situ purification techniques will be used as necessary. By loading
the 780 tonnes of LAB scintillator with O(tonne) of Nd, an experiment
to search for neutrinoless double beta decay can be performed down to
an effective neutrino mass of 100-200~meV at 90\% C.L. with a 3 year
run. SNO+ is fully funded and will take water-fill data in 2012 with
scintillator data following in 2013.

\section{References}
%%%%%%%%%%%%%%%%%%%%%%%%%%%%%%%%%%%%%%%%%%%

\smallskip

%\begin{figure}[h]
%\begin{minipage}{14pc}
%\includegraphics[width=14pc]{name.eps}
%\caption{\label{label}Figure caption for first of two sided figures.}
%\end{minipage}\hspace{2pc}%
%\begin{minipage}{14pc}
%\includegraphics[width=14pc]{name.eps}
%\caption{\label{label}Figure caption for second of two sided figures.}
%\end{minipage} 
%\end{figure}

%\begin{figure}[h]
%\includegraphics[width=14pc]{name.eps}\hspace{2pc}%
%\begin{minipage}[b]{14pc}\caption{\label{label}Figure caption for a narrow figure where the caption is put at the side of the figure.}
%\end{minipage}
%\end{figure}


\begin{thebibliography}{9}

\bibitem{Chen:2005yi}
  M.~C.~Chen,
  %``The SNO liquid scintillator project,''
  Nucl.\ Phys.\ Proc.\ Suppl.\  {\bf 145} (2005) 65.
  %%CITATION = NUPHZ,145,65;%%

%\cite{Mitsui:2011zz}
\bibitem{Mitsui:2011zz}
  T.~Mitsui [KamLAND Collaboration],
  %``Low-energy neutrino physics with KamLAND,''
  Nucl.\ Phys.\ Proc.\ Suppl.\  {\bf 217} (2011) 89.
  %%CITATION = NUPHZ,217,89;%%

\bibitem{Boger:1999bb}
  J.~Boger {\it et al.}  [SNO Collaboration],
  %``The Sudbury Neutrino Observatory,''
  Nucl.\ Instrum.\ Meth.\  A {\bf 449} (2000) 172
  [arXiv:nucl-ex/9910016].
  %%CITATION = NUIMA,A449,172;%%

\bibitem{lozza}
  V.~Lozza, these proceedings (TAUP 2011).

\bibitem{ford}
  R.~Ford, M.~Chen, O.~Chkvorets, D.~Hallman and E.~Vazquez-Jauregui,
  %``SNO+ scintillator purification and assay,''
  AIP Conf.\ Proc.\ \ {\bf 1338} (2011) 183.
  %%CITATION = APCPC,1338,183;%%

\bibitem{borexino}
  G.~Alimonti {\it et al.} [Borexino Collaboration],
  %``The liquid handling systems for the Borexino solar neutrino detector,''
  Nucl.\ Instrum.\ Meth.\ A\ {\bf 609} (2009) 58.
  %%CITATION = NUIMA,A609,58;%%

%\cite{760253}
\bibitem{bnlGd}
  M.~Yeh, A.~Garnov and R.~L.~Hahn,
  %``Gadolinium-loaded liquid scintillator for high-precision measurements of antineutrino oscillations and the mixing angle, Theta(13),''
  Nucl.\ Instrum.\ Meth.\ A\ {\bf 578} (2007) 329.
  %%CITATION = NUIMA,A578,329;%%

\bibitem{bnlNdPur}
  S.~Hans, M.~Yeh, J.~B.~Cumming and R.~L.~Hahn,
  AIP Conf.\ Proc.\ {\bf 1338} (2011) 171.

%\cite{Argyriades:2008pr}
\bibitem{Argyriades:2008pr}
  J.~Argyriades {\it et al.}  [NEMO Collaboration],
  %``Measurement of the Double Beta Decay Half-life of Nd-150 and Search for Neutrinoless Decay Modes with the NEMO-3 Detector,''
  Phys.\ Rev.\ C {\bf 80} (2009) 032501
  [arXiv:0810.0248 [hep-ex]].
  %%CITATION = ARXIV:0810.0248;%%

%\cite{Barea:2009zza}
\bibitem{Barea:2009zza}
  J.~Barea and F.~Iachello,
  %``Neutrinoless double-beta decay in the microscopic interacting boson model,''
  Phys.\ Rev.\ C {\bf 79} (2009) 044301.
  %%CITATION = PHRVA,C79,044301;%%

\end{thebibliography}
\end{document}